\documentclass[english,10pt,twocolumn,psfig,times,mathptm,a4paper,conference]{IEEEtran}
\usepackage[T1]{fontenc}
\usepackage[latin9]{inputenc}
\usepackage{float}
\usepackage{amsmath}
\usepackage{amssymb}
\usepackage{graphicx}
\usepackage{esint}

\makeatletter
\@ifundefined{date}{}{\date{}}

\usepackage{setspace}
\usepackage{pstricks, pst-node, pst-plot, pst-circ}
\usepackage{moredefs}

\usepackage{psfrag}

\ifCLASSINFOpdf
\else
\fi
\newcommand{\HP}{\text{H\hspace{-0.25mm}P}}
\newcommand{\LP}{\text{L\hspace{-0.25mm}P}}

\makeatother

\usepackage{babel}
\begin{document}

\author{\IEEEauthorblockN{Wolfgang Schnurrer, Jürgen Seiler, Michael Schöberl,
and André Kaup} \IEEEauthorblockA{\\
Multimedia Communications and Signal Processing\\
University of Erlangen-Nuremberg, Cauerstr. 7, 91058 Erlangen, Germany\\
Email: \{schnurrer, seiler, schoeberl, kaup\}@lnt.de}}

\title{On the Influence of Clipping in Lossless Predictive and Wavelet Coding
of Noisy Images}

\maketitle


\begin{abstract}
Especially in lossless image coding the obtainable compression ratio
strongly depends on the amount of noise included in the data as all
noise has to be coded, too. Different approaches exist for lossless
image coding. We analyze the compression performance of three kinds
of approaches, namely direct entropy, predictive and wavelet-based
coding. The results from our theoretical model are compared to simulated
results from standard algorithms that base on the three approaches.
As long as no clipping occurs with increasing noise more bits are
needed for lossless compression. We will show that for very noisy
signals it is more advantageous to directly use an entropy coder without
advanced preprocessing steps.
\end{abstract}


\IEEEpeerreviewmaketitle

\section{Introduction}


Lossless compression is an important task in all areas where any modification
of information is not allowed or at least not acceptable. Examples
are among many others measurements for quality assurance, archiving,
surveillance, conservation of evidence material or medical data. Noise
that is contained in the data has also to be coded in this case. In
the medical environment lossy compression is often not acceptable
as the correct diagnosis cannot be guaranteed for the lossy coded
images. But medical images contain a lot of noise because on the one
hand radiation has to be kept low to reduce the risks for the patients
and on the other hand the acquisition time is kept short to avoid
motion artifacts.

Several different approaches exist for lossless coding. We observed
that the performance of the different approaches varies significantly
when the data contains different amounts of noise. Prediction-based
methods like lossless~JPEG \cite{losslessJPEG,wallace1992} and wavelet-based
methods like JPEG~2000 \cite{jpeg2k,christopoulos2002} are advantageous
to code the structural information but become less effective when
the images contain a lot of noise. We will provide a theoretical analysis
on the behavior of the energy of the noise when a wavelet transform
is applied. We will compare this to \emph{direct} entropy coding and
a \emph{predictive} coding scheme.

\begin{figure}[H]
\psfragscanon
\psfrag{sigma}{$\sigma$}
\psfrag{so}{$f[k]$}
\psfrag{si}{} 
\psfrag{n}{$n[k]$}
\psfrag{s}{$s[k]$}
\psfrag{Q}{$Q$}
\psfrag{H}{$H$}
\psfrag{plus}{$+$}
\psfrag{noise}{noise}
\psfrag{dir}{\emph{direct}}
\psfrag{wavelet}{\emph{wavelet}}
\psfrag{pr}{\emph{predictive}}

\includegraphics[width=1\columnwidth]{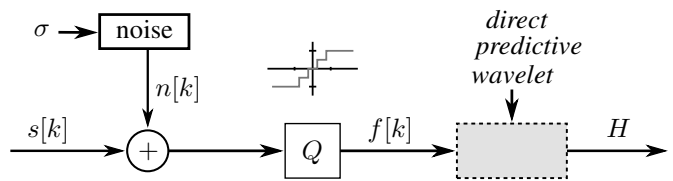}

\psfragscanoff

\vspace*{-2mm}

\protect\caption{\label{fig:SignalModel}Signal model}
\end{figure}

Figure~\ref{fig:SignalModel} shows a block diagram of our signal
model. Noise $n\left[k\right]$ with a standard deviation $\sigma$
is added to the signal $s\left[k\right]$ that contains the structural
information. The signal $f\left[k\right]$ results from quantizing
the noisy signal. One of the methods, i.e., \emph{direct}, \emph{predictive}
and \emph{wavelet}, is then applied in the gray box and the output
is analyzed. In our study we assume that clipping occurs mainly when
the additive noise leads to values that exceed the limits of the quantizer
$Q$.

\vspace*{-6mm}
\begin{figure}[H]
\begin{minipage}[t]{0.3\columnwidth}%
\includegraphics[width=1\columnwidth]{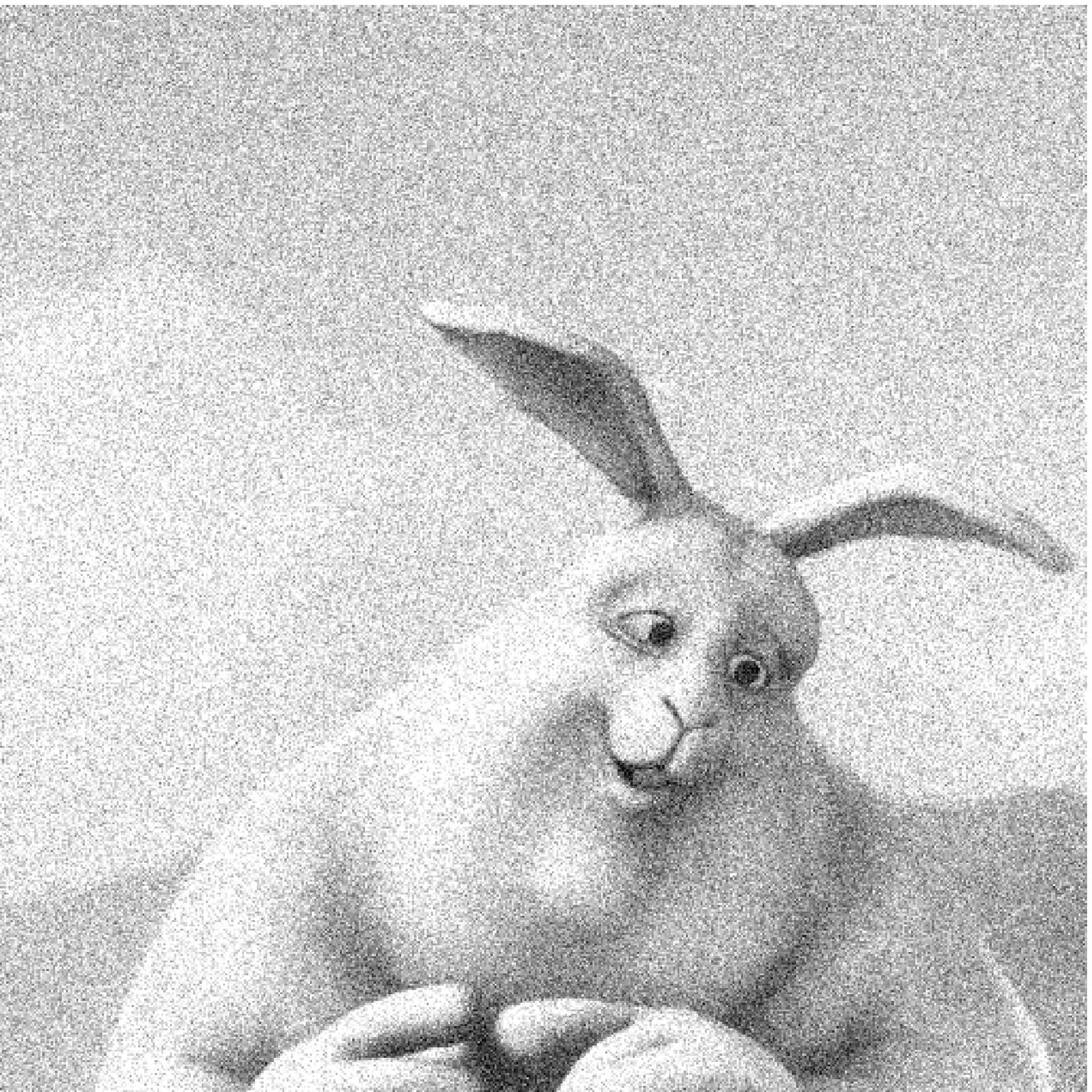}

\vspace*{0.4mm}\includegraphics[width=1\columnwidth]{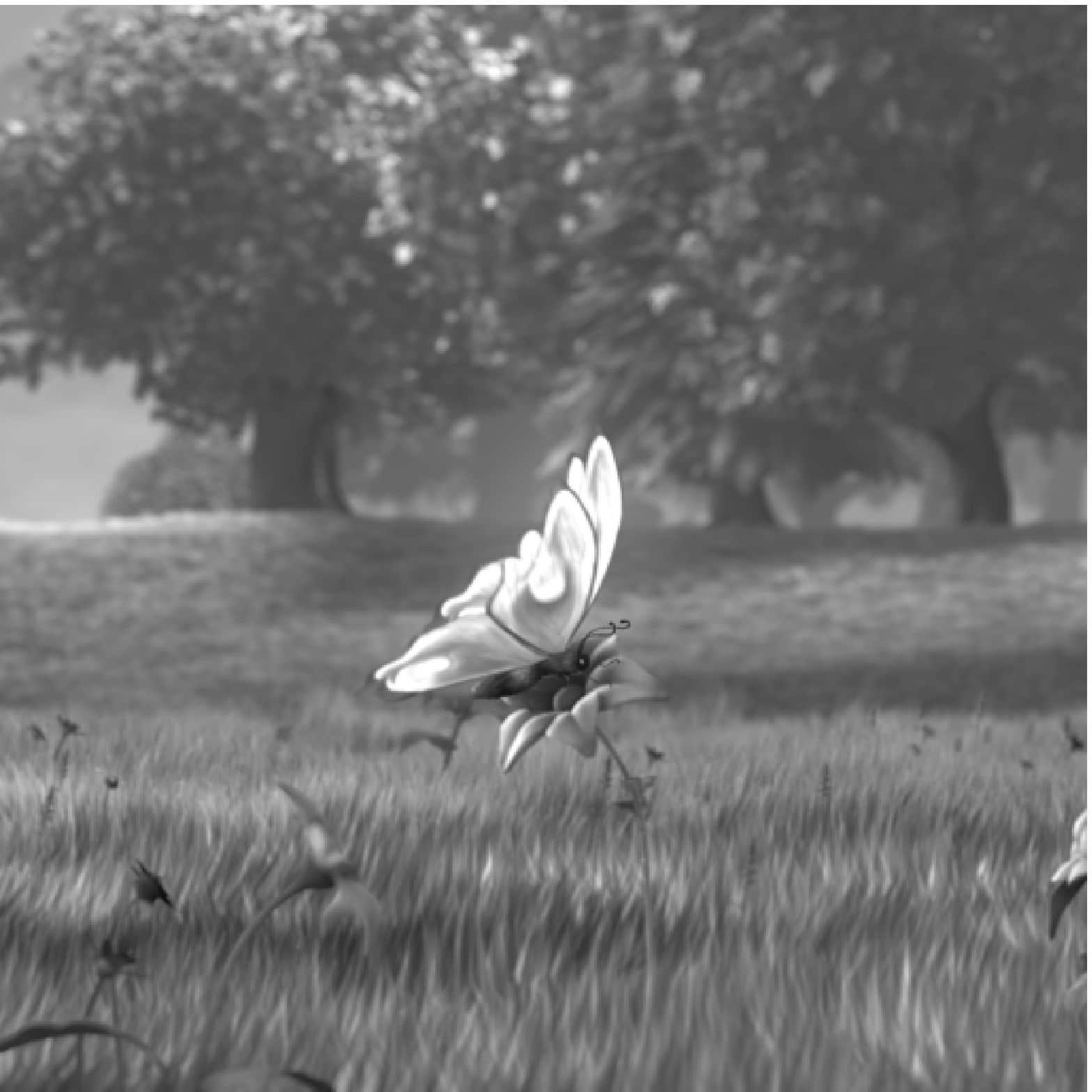}%
\end{minipage}~%
\begin{minipage}[t]{0.6\columnwidth}%
\vspace{-1.3cm}
\input{fig/hist_n30_3on2_file2.tex}%
\end{minipage}

\vspace*{1.5mm}

\hspace{0.06\columnwidth}(a) $\uparrow$ / (b) $\downarrow$\hspace{0.4\columnwidth}(c)

\vspace*{-1.5mm}

\protect\caption{\label{fig:SimulationTestimages}Two detail images from the \emph{Big
Buck Bunny} sequence (a), (b). (a) is shown with additive noise with
$\sigma=30$, (b) is shown original. (c) shows histograms of the image
(a) with additive noise (solid red) and original of image (a), i.e.,
without additive noise (dashed blue)}
\end{figure}

In Section~2 we present the analysis. The description of our simulation
and results are given in Section~3. Section~4 will conclude this
study.

\section{Theoretical Analysis of Noise Impact}

We compare three different approaches for coding a signal without
loss. The first method is called \emph{direct} as the samples are
directly entropy coded without any preprocessing. The second method
is called \emph{predictive}. Before entropy coding it is possible
to subtract a prediction where the predictor for the current sample
is computed from already decoded samples. The various predictors differ
by the number and the weight of the incorporated samples.  We analyze
the prediction from one previous sample. The combination of more samples
leads to a noise variance reduction due to averaging. But the overall
noise variance of the predictor will stay greater than zero. The third
class of methods in our analysis is based on the wavelet transform.
Instead of subtracting a prediction, the samples are transformed and
the coefficients from the sub-bands are then entropy coded. In our
analysis we compare two different wavelets, the \emph{Haar wavelet}
and the \emph{LeGall~5/3 wavelet}.

We assume that the input signal $f\left[k\right]$ consists of the
structural information $s\left[k\right]$ with additive noise $n\left[k\right]$
after quantization as shown in Figure~\ref{fig:SignalModel}. For
simplicity we show the analysis for the one dimensional case only.
At first we neglect the structural part and quantization step. We
analyze and compare the output of the different methods. We then show
how to calculate the entropy and finally add the structural information
of the signal and the quantization step to our modeling.

\subsection{Noise Variance for Different Coding Methods}

For our analysis we are mainly interested in the noise part. We consider
the structural signal $s\left[k\right]=0$ in this subsection and
assume Gaussian noise $n\left[k\right]$ with zero mean and a variance
$\sigma^{2}$. We analyze the influence of the different methods
on the noise by comparing the noise variance at their output.

The \emph{direct} method does not apply preprocessing and so the error
distribution does not change and stays equal to 
\begin{equation}
\mathcal{E}\left\{ n_{\text{direct}}^{2}[k]\right\} =\sigma^{2}.
\end{equation}

The second method based on prediction uses one previous sample for
the calculation of the predictor $p[k]=f[k-1]$ and will be called
\emph{predictive}. The residuum of the predictor $r[k]=f[k]-p\left[k\right]$
is then coded and the variance of the noise in the resulting sample
doubles to 
\begin{equation}
\mathcal{E}\left\{ n_{p}^{2}[k]\right\} =\mathcal{E}\left\{ \left(n[k]-n[k-1]\right)^{2}\right\} =\sigma^{2}+\sigma^{2}=2\sigma^{2}.\label{eq:varNoisePrediction1}
\end{equation}

Wavelet transforms consist of a filter pair for the computation of
the high $\HP$- and the low $\LP$-band. For the \emph{Haar wavelet,}
the samples of the $\HP_{\text{Haar}}$- and the $\LP_{\text{Haar}}$-band
are calculated to $\HP_{\text{Haar}}\left[k\right]=f\left[2k\right]-f\left[2k-1\right]$
and $\LP_{\text{Haar}}\left[k\right]=\left\lfloor \frac{1}{2}\left(f\left[2k\right]-f\left[2k-1\right]\right)\right\rfloor $.
The rounding operation is due to the integer wavelet transform \cite{calderbank1997}
for lossless coding. The error variance for the coefficients of the
$\HP_{\text{Haar}}$ band calculates to 
\begin{equation}
\mathcal{E}\left\{ n_{\HP_{\text{Haar}}}^{2}[k]\right\} =\mathcal{E}\left\{ \left(n[2k]-n[2k-1]\right)^{2}\right\} =2\sigma^{2}.
\end{equation}
For the $\LP_{\text{Haar}}$-band the error variance is equal to 
\begin{equation}
\mathcal{E}\left\{ n_{\LP_{\text{Haar}}}^{2}[k]\right\} =\frac{1}{4}\sigma^{2}+\frac{1}{4}\sigma^{2}=\frac{1}{2}\sigma^{2}.
\end{equation}
For the \emph{LeGall~5/3 wavelet} the samples of the $\HP_{\text{LeGall}}$-
and the $\LP_{\text{LeGall}}$- band are calculated to 
\begin{equation}
\HP_{\text{LeGall}}\left[k\right]=f\left[2k\right]-\left\lfloor \frac{1}{2}\left(f\left[2k-1\right]+f\left[2k+1\right]\right)\right\rfloor \label{eq:H53}
\end{equation}
and 
\begin{equation}
\LP_{\text{LeGall}}\hspace{-0.5mm}\left[k\right]\hspace{-1mm}=\hspace{-1mm}f\hspace{-0.5mm}\left[2k\hspace{-0.5mm}-\hspace{-0.5mm}1\right]\hspace{-1mm}+\hspace{-1.5mm}\left\lfloor \frac{1}{4}\hspace{-0.5mm}\left(\HP_{\text{LeGall}}\hspace{-0.5mm}\left[k\right]\hspace{-0.5mm}+\hspace{-0.5mm}\HP_{\text{LeGall}}\hspace{-0.5mm}\left[k\hspace{-0.5mm}-\hspace{-0.5mm}1\right]\right)\hspace{-0.5mm}\right\rfloor \label{eq:L53}
\end{equation}
as given in \cite{calderbank1997}. The error variance for the coefficients
of the $\HP_{\text{LeGall}}$-band calculates to 
\begin{equation}
\mathcal{E}\left\{ n_{\HP_{\text{LeGall}}}^{2}[k]\right\} =\frac{1}{4}\sigma^{2}+\sigma^{2}+\frac{1}{4}\sigma^{2}=\frac{3}{2}\sigma^{2}.
\end{equation}
For the $\LP_{\text{LeGall}}$-band the error variance needs to be
calculated from the filter representation of the wavelet to preserve
the signal independence. The error variance is then equal to 
\begin{equation}
\mathcal{E}\left\{ n_{\LP_{\text{LeGall}}}^{2}[k]\right\} =\frac{1^{2}+2^{2}+6^{2}+2^{2}+1^{2}}{64}\sigma^{2}=\frac{46}{64}\sigma^{2}.
\end{equation}

The combined error variance cannot be computed by averaging the two
values $\mathcal{E}\left\{ n_{\LP}^{2}\right\} $ and $\mathcal{E}\left\{ n_{\HP}^{2}\right\} $
from the subbands. The two subbands are calculated from the same signal
values so the independence assumption does not hold anymore. In the
next subsection we will show how the values can be combined.

\subsection{From Noise Variance to Bits}

For the \emph{wavelet} approach a combined variance cannot be computed
by simply averaging the variances from the \HP{}- and \LP{}-bands.
In order to get an estimation for the number of bits needed for coding
the signal processed by the different methods with an optimum entropy
coder we use the entropy 
\begin{equation}
H=-\sum_{i}p_{i}\text{log}_{2}\left(p_{i}\right)\ \left[\text{bit per sample}\right].\label{eq:entropy}
\end{equation}
 The probabilities $p_{i}$ result from integrals over the probability
density function with 
\begin{equation}
p_{i}=\int_{i-\frac{1}{2}}^{i+\frac{1}{2}}\frac{1}{\sqrt{2\pi}\sigma^{2}}e^{-\frac{1}{2}\left(\frac{x}{\sigma}\right)^{2}}\,\text{d}x.\label{eq:piWithoutClipping}
\end{equation}
Without considering clipping on the upper and lower end of the co-domain
$i\in\mathbb{Z}$ holds. For this case \eqref{eq:entropy} can be
solved analytically and be approximated by a formula that is dependent
on the standard deviation $\sigma$ of the noise 
\begin{equation}
H_{\text{direct}}\left(\sigma\right)\approx\log_{2}\left(\sigma\sqrt{2\pi e}\right).\label{eq:gaussEntropy}
\end{equation}

Now we can calculate the entropy for the different methods by evaluating
\eqref{eq:gaussEntropy} for the variances derived in the previous
subsection. 

For the \emph{predictive} scheme the resulting entropy is given by
\begin{equation}
H_{\text{predictive}}\left(\sigma\right)\approx H(\sqrt{2}\sigma)=\frac{1}{2}\,\text{bit}+H_{\text{direct}}\left(\sigma\right).\label{eq:entropyPrediction1}
\end{equation}

After the wavelet transform, the signal is represented by coefficients
in the high and the low band in a ratio of~1:1. So the overall entropy
can be computed by summing up the entropy of the high and the low
band containing half the samples each. For the \emph{Haar wavelet}
this results in
\begin{eqnarray}
H_{\text{Haar}}\left(\sigma\right) & \approx & \frac{1}{2}\left(H\left(\sqrt{\frac{1}{2}}\sigma\right)+H\left(\sqrt{2}\sigma\right)\right)\label{eq:entropyHaar}\\
 & = & \log_{2}\left(\sigma\sqrt{2\pi e}\right)=H_{\text{direct}}\left(\sigma\right)\nonumber 
\end{eqnarray}
and for the \emph{LeGall 5/3 wavelet} in 
\begin{eqnarray}
H_{\text{LeGall}}\left(\sigma\right) & \approx & \frac{1}{2}\left(H\left(\sqrt{\frac{3}{2}}\sigma\right)+H\left(\sqrt{\frac{46}{64}}\sigma\right)\right)\nonumber \\
 & = & 0.027\,\text{bit}+H_{\text{direct}}\left(\sigma\right).\label{eq:entropyL53}
\end{eqnarray}

Even though the value range doubles for the $H_{\text{Haar}}$-band
that contains half of all coefficients, the entropy of the whole signal
stays the same for the \emph{Haar wavelet}. Comparing the two wavelets,
the entropy increases slightly for the \emph{LeGall~5/3 wavelet}
by an offset of $0.027$~bit per sample. 

Comparing the entropy of the different methods we can conclude that
without clipping all the methods need the same rate for coding the
noise up to an offset that does not depend on $\sigma$. While this
offset is very small for the \emph{wavelet}-based methods for the
\emph{predictive} scheme that uses one previous sample the coding
of the noise is more expensive by 0.5~bit per sample.

\subsection{The Influence of Clipping}

We now extend the modeling by the structural information and the quantization.
As we are interested in lossless coding we are limited to integer
values in order to avoid rounding errors. To analyze the impact of
clipping to zero mean noise we use a structural signal $s\left[k\right]$
with a constant value in the center of the co-domain $\mu=\left\lceil \frac{2^{8}-1}{2}\right\rceil $.
For this we extend \eqref{eq:entropy} by a limitation of $i$. The
probabilities $p_{i}$, $i\in0..255$ for a co-domain of 8~bit, of
the signal result from integrating of the probability density function
$f_{N}\left(n\right)$ of the noise over the bin size. The calculation
of the probabilities $p_{i}$ is given by
\begin{equation}
p_{i}=\begin{cases}
\int_{-\infty}^{-127.5}\ f_{N}\left(\nu\right)\,\text{d}\nu & \text{for }i=0\\
\int_{126.5}^{\infty}\ f_{N}\left(\nu\right)\,\text{d}\nu & \text{for }i=255\\
\int_{i-0.5}^{i+0.5}\ f_{N}\left(\nu\right)\,\text{d}\nu & \text{for }0<i<254
\end{cases}.\label{eq:piCalc}
\end{equation}

Values outside the co-domain are clipped to the minimum~0 and maximum~255
code value respectively. This leads to peaks in the probability distribution
of the signal at the borders of the co-domain as in Figure~\ref{fig:SimulationTestimages}~(c).

In our analysis we assume statistically independent Gaussian noise,
so the probability density function of the noise equals $f_{N}\left(n\right)=\frac{1}{\sqrt{2\pi}\sigma^{2}}e^{-\frac{1}{2}\left(\frac{x}{\sigma}\right)^{2}}$.
As we assume a signal with a constant code value only, the probability
distribution of the signal equals $p_{i}$. The probability distribution
of the input signal can now be combined according to the presented
methods. 

Then the entropy can be calculated by inserting the values of the
probabilities from \ref{eq:piCalc} into the entropy formula~\eqref{eq:entropy}.
For the \emph{wavelet}-based methods the entropy for the \HP{}-band
and for the \LP{}-band have to be calculated separately. The resulting
values are then summed up by taking into account that each band contains
half of the samples.

\subsection{Theoretical Results}

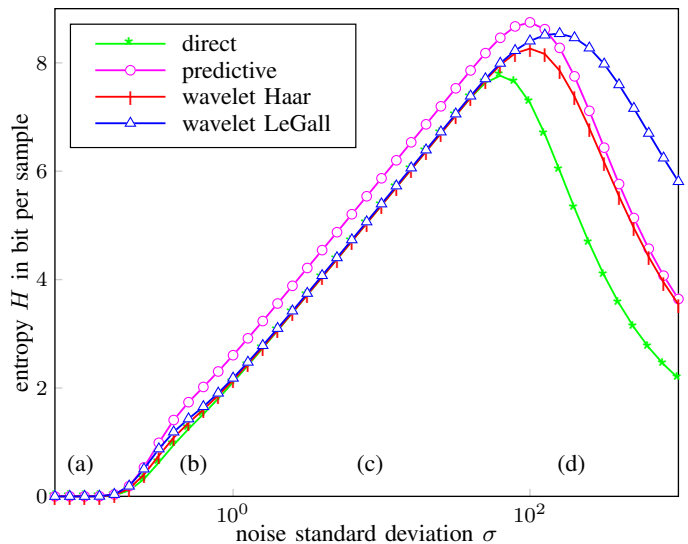
\begin{figure}
%
%
%
\providelength{\AxesLineWidth}       \setlength{\AxesLineWidth}{0.5pt}%
\providelength{\plotwidth}           \setlength{\plotwidth}{8.2cm}
\providelength{\LineWidth}           \setlength{\LineWidth}{0.7pt}%
\providelength{\MarkerSize}          \setlength{\MarkerSize}{4pt}%
\newrgbcolor{GridColor}{0.8 0.8 0.8}%
%
\psset{xunit=0.238095\plotwidth,yunit=0.087634\plotwidth}%
\begin{pspicture}(-1.507317,-0.974113)(3.005122,9.013916)%


\psline[linewidth=\AxesLineWidth,linecolor=GridColor](0.000000,0.000000)(0.000000,0.136933)
\psline[linewidth=\AxesLineWidth,linecolor=GridColor](2.000000,0.000000)(2.000000,0.136933)
\psline[linewidth=\AxesLineWidth,linecolor=GridColor](-1.200000,0.000000)(-1.149600,0.000000)
\psline[linewidth=\AxesLineWidth,linecolor=GridColor](-1.200000,2.000000)(-1.149600,2.000000)
\psline[linewidth=\AxesLineWidth,linecolor=GridColor](-1.200000,4.000000)(-1.149600,4.000000)
\psline[linewidth=\AxesLineWidth,linecolor=GridColor](-1.200000,6.000000)(-1.149600,6.000000)
\psline[linewidth=\AxesLineWidth,linecolor=GridColor](-1.200000,8.000000)(-1.149600,8.000000)


{ \footnotesize 
\rput[t](0.000000,-0.136933){$10^{0}$}
\rput[t](2.000000,-0.136933){$10^{2}$}
\rput[r](-1.250400,0.000000){$0$}
\rput[r](-1.250400,2.000000){$2$}
\rput[r](-1.250400,4.000000){$4$}
\rput[r](-1.250400,6.000000){$6$}
\rput[r](-1.250400,8.000000){$8$}
} 

\psframe[linewidth=\AxesLineWidth,dimen=middle](-1.200000,0.000000)(3.000000,9.000000)

{ \small 
\rput[b](0.900000,-0.974113){
\begin{tabular}{c}
noise standard deviation $\sigma$\\
\end{tabular}
}

\rput[t]{90}(-1.507317,4.500000){
\begin{tabular}{c}
entropy $H$ in bit per sample\\
\end{tabular}
}
} 

\newrgbcolor{color328.0016}{0  1  0}
\psline[plotstyle=line,linejoin=1,showpoints=true,dotstyle=Basterisk,dotsize=\MarkerSize,linestyle=solid,linewidth=\LineWidth,linecolor=color328.0016]
(-1.200000,0.000000)(-1.100000,0.000000)(-1.000000,0.000013)(-0.900000,0.001158)(-0.800000,0.018830)
(-0.700000,0.107342)(-0.600000,0.318010)(-0.500000,0.625292)(-0.400000,0.950901)(-0.300000,1.244059)
(-0.200000,1.518985)(-0.100000,1.804370)(0.000000,2.104833)(0.100000,2.416253)(0.200000,2.735024)
(0.300000,3.058618)(0.400000,3.385332)(0.500000,3.714046)(0.600000,4.044035)(0.700000,4.374834)
(0.800000,4.706146)(0.900000,5.037783)(1.000000,5.369625)(1.100000,5.701596)(1.200000,6.033649)
(1.300000,6.365753)(1.400000,6.697889)(1.500000,7.029794)(1.600000,7.355289)(1.700000,7.633020)
(1.800000,7.762310)(1.900000,7.644008)(2.000000,7.262896)(2.100000,6.686759)(2.200000,6.011957)
(2.300000,5.320544)(2.400000,4.665978)(2.500000,4.076036)(2.600000,3.560812)(2.700000,3.120006)
(2.800000,2.748004)(2.900000,2.436980)(3.000000,2.178618)

\newrgbcolor{color329.0011}{1  0  1}
\psline[plotstyle=line,linejoin=1,showpoints=true,dotstyle=Bo,dotsize=\MarkerSize,linestyle=solid,linewidth=\LineWidth,linecolor=color329.0011]
(-1.200000,0.000000)(-1.100000,0.000000)(-1.000000,0.000025)(-0.900000,0.002172)(-0.800000,0.034425)
(-0.700000,0.189354)(-0.600000,0.534125)(-0.500000,0.988045)(-0.400000,1.408825)(-0.300000,1.737536)
(-0.200000,2.018660)(-0.100000,2.304373)(0.000000,2.604834)(0.100000,2.916253)(0.200000,3.235024)
(0.300000,3.558618)(0.400000,3.885332)(0.500000,4.214046)(0.600000,4.544035)(0.700000,4.874834)
(0.800000,5.206146)(0.900000,5.537783)(1.000000,5.869625)(1.100000,6.201596)(1.200000,6.533649)
(1.300000,6.865753)(1.400000,7.197890)(1.500000,7.529971)(1.600000,7.860354)(1.700000,8.179200)
(1.800000,8.462569)(1.900000,8.672351)(2.000000,8.747100)(2.100000,8.621285)(2.200000,8.276255)
(2.300000,7.751037)(2.400000,7.114310)(2.500000,6.434426)(2.600000,5.763943)(2.700000,5.136628)
(2.800000,4.570494)(2.900000,4.072375)(3.000000,3.642029)

\newrgbcolor{color330.0011}{1  0  0}
\psline[plotstyle=line,linejoin=1,showpoints=true,dotstyle=B|,dotsize=\MarkerSize,linestyle=solid,linewidth=\LineWidth,linecolor=color330.0011]
(-1.200000,0.000000)(-1.100000,0.000000)(-1.000000,0.000019)(-0.900000,0.001629)(-0.800000,0.025822)
(-0.700000,0.142188)(-0.600000,0.402552)(-0.500000,0.750475)(-0.400000,1.082881)(-0.300000,1.355575)
(-0.200000,1.601977)(-0.100000,1.862500)(0.000000,2.144210)(0.100000,2.442312)(0.200000,2.751990)
(0.300000,3.069543)(0.400000,3.392316)(0.500000,3.718490)(0.600000,4.046854)(0.700000,4.376619)
(0.800000,4.707275)(0.900000,5.038496)(1.000000,5.370075)(1.100000,5.701880)(1.200000,6.033828)
(1.300000,6.365866)(1.400000,6.697961)(1.500000,7.030017)(1.600000,7.360383)(1.700000,7.679244)
(1.800000,7.963024)(1.900000,8.176002)(2.000000,8.263062)(2.100000,8.163506)(2.200000,7.857769)
(2.300000,7.380210)(2.400000,6.794088)(2.500000,6.163427)(2.600000,5.538085)(2.700000,4.950535)
(2.800000,4.418466)(2.900000,3.948965)(3.000000,3.542326)

\newrgbcolor{color331.0011}{0  0  1}
\psline[plotstyle=line,linejoin=1,showpoints=true,dotstyle=Btriangle,dotsize=\MarkerSize,linestyle=solid,linewidth=\LineWidth,linecolor=color331.0011]
(-1.200000,0.000000)(-1.100000,0.000000)(-1.000000,0.000028)(-0.900000,0.002312)(-0.800000,0.035734)
(-0.700000,0.189245)(-0.600000,0.505043)(-0.500000,0.876528)(-0.400000,1.188488)(-0.300000,1.430572)
(-0.200000,1.656217)(-0.100000,1.905587)(0.000000,2.181553)(0.100000,2.476127)(0.200000,2.783422)
(0.300000,3.099416)(0.400000,3.421188)(0.500000,3.746724)(0.600000,4.074683)(0.700000,4.404191)
(0.800000,4.734684)(0.900000,5.065803)(1.000000,5.397317)(1.100000,5.729081)(1.200000,6.061003)
(1.300000,6.393025)(1.400000,6.725110)(1.500000,7.057158)(1.600000,7.387495)(1.700000,7.706316)
(1.800000,7.992643)(1.900000,8.227497)(2.000000,8.403425)(2.100000,8.513544)(2.200000,8.540243)
(2.300000,8.463937)(2.400000,8.274585)(2.500000,7.978151)(2.600000,7.596434)(2.700000,7.160108)
(2.800000,6.700460)(2.900000,6.243860)(3.000000,5.809536)

{ \small 
\rput[tl](-1.099200,8.726135){%
\psframebox[framesep=0pt,linewidth=\AxesLineWidth]{\psframebox*{\begin{tabular}{l}
\Rnode{a1}{\hspace*{0.0ex}} \hspace*{0.7cm} \Rnode{a2}{~~direct} \\
\Rnode{a3}{\hspace*{0.0ex}} \hspace*{0.7cm} \Rnode{a4}{~~predictive} \\
\Rnode{a5}{\hspace*{0.0ex}} \hspace*{0.7cm} \Rnode{a6}{~~wavelet Haar} \\
\Rnode{a7}{\hspace*{0.0ex}} \hspace*{0.7cm} \Rnode{a8}{~~wavelet LeGall} \\
\end{tabular}}
\ncline[linestyle=solid,linewidth=\LineWidth,linecolor=color328.0016]{a1}{a2} \ncput{\psdot[dotstyle=Basterisk,dotsize=\MarkerSize,linecolor=color328.0016]}
\ncline[linestyle=solid,linewidth=\LineWidth,linecolor=color329.0011]{a3}{a4} \ncput{\psdot[dotstyle=Bo,dotsize=\MarkerSize,linecolor=color329.0011]}
\ncline[linestyle=solid,linewidth=\LineWidth,linecolor=color330.0011]{a5}{a6} \ncput{\psdot[dotstyle=B|,dotsize=\MarkerSize,linecolor=color330.0011]}
\ncline[linestyle=solid,linewidth=\LineWidth,linecolor=color331.0011]{a7}{a8} \ncput{\psdot[dotstyle=Btriangle,dotsize=\MarkerSize,linecolor=color331.0011]}
}%
}%
} 

{ \small 
\newrgbcolor{color311.0011}{0  0  0}
\uput{0pt}[0](2.160000,0.602791){%
\psframebox[framesep=1pt,fillstyle=solid,linestyle=none,linewidth=0.5pt]{\begin{tabular}{@{}c@{}}
(d)\\[-0.3ex]
\end{tabular}}}
\newrgbcolor{color299.0011}{0  0  0}
\uput{0pt}[0](0.805161,0.602791){%
\psframebox[framesep=1pt,fillstyle=solid,linestyle=none,linewidth=0.5pt]{\begin{tabular}{@{}c@{}}
(c)\\[-0.3ex]
\end{tabular}}}
\newrgbcolor{color287.0011}{0  0  0}
\uput{0pt}[0](-0.387097,0.602791){%
\psframebox[framesep=1pt,fillstyle=solid,linestyle=none,linewidth=0.5pt]{\begin{tabular}{@{}c@{}}
(b)\\[-0.3ex]
\end{tabular}}}
\newrgbcolor{color241.0011}{0  0  0}
\uput{0pt}[0](-1.145806,0.602791){%
\psframebox[framesep=1pt,fillstyle=solid,linestyle=none,linewidth=0.5pt]{\begin{tabular}{@{}c@{}}
(a)\\[-0.3ex]
\end{tabular}}}
} 

\end{pspicture}%

\vspace*{-2mm}

\protect\caption{\label{fig:TheoryEntropyOverVariance}Theoretical results - entropy
in bit per sample plotted over the standard deviation $\sigma$ of
the noise on the abscissa in a lin-log plot\vspace*{-1.5mm}}
\end{figure}

The resulting curves are plotted in Figure~\ref{fig:TheoryEntropyOverVariance}.
The results strongly depend on the $\mu$ chosen for the structural
part of the signal. The value of $\mu$ in the center of the co-domain
is the ideal case because for a smaller or greater value of $\mu$
clipping is introduced at smaller values of $\sigma$. The plot can
be divided into four areas along the abscissa denoted by (a)-(d).
In area~(a) the noise is quantized to zero and thus has no influence
on the entropy. When $\sigma$ gets bigger and only a few samples
are affected by noise, that is not quantized to zero, the curves have
a different slope as area~(b) shows. The reason for the steeper
slope of the \emph{predictive} and \emph{wavelet}-based methods is
the combination of several samples. As soon as the noise component
in one sample is above quantization, several coefficients are affected
and cause an increase of the entropy. For bigger values of $\sigma$
more samples are affected by noise. As long as the noise is small
enough such that no clipping is occurring, the curves run in parallel,
as shown in area~(c). The offsets correspond to the derivation in
the previous subsection. Clipping begins to occur for larger values
of the standard deviation~$\sigma$ as shown in area~(d). \emph{Direct}
entropy coding can exploit the rising number of clipped values better
resulting in a lower entropy. The entropy of the \emph{predictive}
scheme and the \emph{wavelet}-based method drops slower due to the
combination of clipped samples with unclipped ones. The \emph{LeGall~5/3
wavelet} is even more sensitive to this than the \emph{Haar wavelet}
because of the greater filter length.

\section{Simulation Results}

In our simulation we used different images and added Gaussian noise
with increasing standard deviation $\sigma$ before coding them with
different algorithms. To validate our model we first used an image
of size 512x512 pixels with a constant code value of 128.

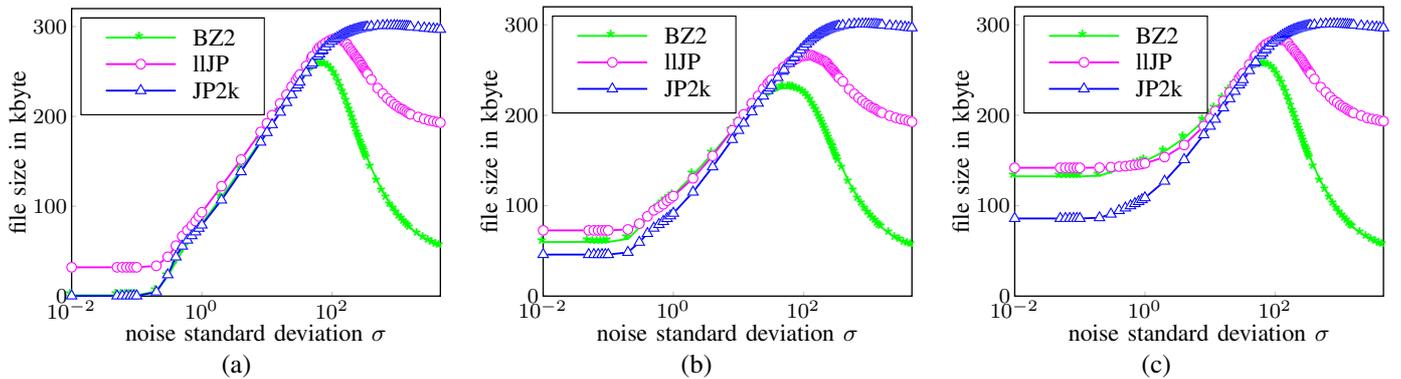
\begin{figure*}[!t]
%
%
%
\providelength{\AxesLineWidth}       \setlength{\AxesLineWidth}{0.5pt}%
\providelength{\plotwidth}           \setlength{\plotwidth}{4.85cm}
\providelength{\LineWidth}           \setlength{\LineWidth}{0.7pt}%
\providelength{\MarkerSize}          \setlength{\MarkerSize}{4pt}%
\newrgbcolor{GridColor}{0.8 0.8 0.8}%
%
\psset{xunit=0.176592\plotwidth,yunit=0.002459\plotwidth}%
\begin{pspicture}(-2.992442,-58.699153)(3.674434,320.838559)%


\psline[linewidth=\AxesLineWidth,linecolor=GridColor](-2.000000,0.000000)(-2.000000,4.880415)
\psline[linewidth=\AxesLineWidth,linecolor=GridColor](0.000000,0.000000)(0.000000,4.880415)
\psline[linewidth=\AxesLineWidth,linecolor=GridColor](2.000000,0.000000)(2.000000,4.880415)
\psline[linewidth=\AxesLineWidth,linecolor=GridColor](-2.000000,0.000000)(-1.932047,0.000000)
\psline[linewidth=\AxesLineWidth,linecolor=GridColor](-2.000000,100.000000)(-1.932047,100.000000)
\psline[linewidth=\AxesLineWidth,linecolor=GridColor](-2.000000,200.000000)(-1.932047,200.000000)
\psline[linewidth=\AxesLineWidth,linecolor=GridColor](-2.000000,300.000000)(-1.932047,300.000000)


{ \footnotesize 
\rput[t](-2.000000,-4.880415){$10^{-2}$}
\rput[t](0.000000,-4.880415){$10^{0}$}
\rput[t](2.000000,-4.880415){$10^{2}$}
\rput[r](-2.067953,0.000000){$0$}
\rput[r](-2.067953,100.000000){$100$}
\rput[r](-2.067953,200.000000){$200$}
\rput[r](-2.067953,300.000000){$300$}
} 

\psframe[linewidth=\AxesLineWidth,dimen=middle](-2.000000,0.000000)(3.662758,320.000000)

{ \small 
\rput[b](0.831379,-58.699153){
\begin{tabular}{c}
noise standard deviation $\sigma$\\
\end{tabular}
}

\rput[t]{90}(-2.992442,160.000000){
\begin{tabular}{c}
file size in kbyte\\
\end{tabular}
}
} 

\newrgbcolor{color593.0037}{0  1  0}
\psline[plotstyle=line,linejoin=1,showpoints=true,dotstyle=Basterisk,dotsize=\MarkerSize,linestyle=solid,linewidth=\LineWidth,linecolor=color593.0037]
(-2.000000,0.944336)(-1.301030,0.951172)(-1.221849,0.949219)(-1.154902,0.949219)(-1.096910,0.940430)
(-1.045757,0.971680)(-1.000000,0.946289)(-0.698970,5.391602)(-0.522879,21.887695)(-0.397940,38.044922)
(-0.301030,50.167969)(-0.221849,58.704102)(-0.154902,65.998047)(-0.096910,71.610352)(-0.045757,76.360352)
(0.000000,80.834961)(0.301030,109.991211)(0.602060,141.130859)(0.903090,173.078125)(1.000000,184.057617)
(1.079181,192.454102)(1.204120,205.153320)(1.301030,215.586914)(1.380211,224.338867)(1.447158,231.284180)
(1.477121,234.385742)(1.505150,237.123047)(1.602060,246.966797)(1.681241,253.003906)(1.698970,254.041992)
(1.778151,256.808594)(1.806180,257.248047)(1.845098,257.490234)(1.857332,257.484375)(1.903090,256.868164)
(1.954243,254.943359)(2.000000,251.753906)(2.041393,247.750000)(2.079181,242.852539)(2.113943,237.750000)
(2.146128,231.983398)(2.176091,226.457031)(2.204120,221.243164)(2.230449,215.909180)(2.255273,210.974609)
(2.278754,206.450195)(2.301030,201.717773)(2.322219,197.264648)(2.342423,192.953125)(2.361728,188.747070)
(2.380211,185.072266)(2.397940,181.028320)(2.414973,177.935547)(2.431364,174.900391)(2.447158,171.100586)
(2.462398,168.228516)(2.477121,165.326172)(2.491362,162.574219)(2.505150,160.227539)(2.518514,157.536133)
(2.531479,155.301758)(2.544068,153.059570)(2.602060,142.450195)(2.698970,127.253906)(2.778151,116.491211)
(2.845098,108.095703)(2.903090,101.125977)(2.954243,96.358398)(3.000000,92.294922)(3.041393,87.908203)
(3.079181,85.007812)(3.113943,82.187500)(3.146128,79.684570)(3.176091,77.368164)(3.204120,75.780273)
(3.322219,68.951172)(3.414973,64.085938)(3.491362,61.350586)(3.556303,58.453125)(3.612784,56.767578)
(3.662758,55.443359)

\newrgbcolor{color594.0027}{1  0  1}
\psline[plotstyle=line,linejoin=1,showpoints=true,dotstyle=Bo,dotsize=\MarkerSize,linestyle=solid,linewidth=\LineWidth,linecolor=color594.0027]
(-2.000000,32.065430)(-1.301030,32.065430)(-1.221849,32.065430)(-1.154902,32.065430)(-1.096910,32.065430)
(-1.045757,32.065430)(-1.000000,32.065430)(-0.698970,33.685547)(-0.522879,43.743164)(-0.397940,56.158203)
(-0.301030,66.653320)(-0.221849,73.245117)(-0.154902,78.254883)(-0.096910,83.349609)(-0.045757,88.683594)
(0.000000,93.294922)(0.301030,122.317383)(0.602060,151.966797)(0.903090,182.729492)(1.000000,192.296875)
(1.079181,201.373047)(1.204120,214.438477)(1.301030,224.170898)(1.380211,233.234375)(1.447158,240.490234)
(1.477121,243.786133)(1.505150,246.521484)(1.602060,256.145508)(1.681241,265.042969)(1.698970,267.068359)
(1.778151,274.215820)(1.806180,276.800781)(1.845098,278.834961)(1.857332,279.493164)(1.903090,281.996094)
(1.954243,284.214844)(2.000000,285.757812)(2.041393,286.471680)(2.079181,286.605469)(2.113943,286.766602)
(2.146128,285.717773)(2.176091,283.521484)(2.204120,281.833984)(2.230449,280.196289)(2.255273,276.750977)
(2.278754,274.828125)(2.301030,272.371094)(2.322219,270.423828)(2.342423,268.352539)(2.361728,266.365234)
(2.380211,264.936523)(2.397940,263.467773)(2.414973,261.506836)(2.431364,260.028320)(2.447158,257.662109)
(2.462398,256.392578)(2.477121,254.173828)(2.491362,252.731445)(2.505150,251.485352)(2.518514,249.789062)
(2.531479,248.428711)(2.544068,247.234375)(2.602060,240.903320)(2.698970,231.542969)(2.778151,225.429688)
(2.845098,220.291992)(2.903090,216.574219)(2.954243,213.313477)(3.000000,211.226562)(3.041393,209.023438)
(3.079181,207.896484)(3.113943,206.186523)(3.146128,204.809570)(3.176091,203.378906)(3.204120,202.329102)
(3.322219,199.333984)(3.414973,197.132812)(3.491362,195.482422)(3.556303,194.708008)(3.612784,193.759766)
(3.662758,193.083984)

\newrgbcolor{color595.0027}{0  0  1}
\psline[plotstyle=line,linejoin=1,showpoints=true,dotstyle=Btriangle,dotsize=\MarkerSize,linestyle=solid,linewidth=\LineWidth,linecolor=color595.0027]
(-2.000000,0.178711)(-1.301030,0.178711)(-1.221849,0.178711)(-1.154902,0.178711)(-1.096910,0.178711)
(-1.045757,0.178711)(-1.000000,0.178711)(-0.698970,4.553711)(-0.522879,23.743164)(-0.397940,43.518555)
(-0.301030,55.629883)(-0.221849,62.287109)(-0.154902,67.236328)(-0.096910,71.458008)(-0.045757,75.471680)
(0.000000,79.105469)(0.301030,106.912109)(0.602060,138.385742)(0.903090,171.369141)(1.000000,182.016602)
(1.079181,190.830078)(1.204120,204.687500)(1.301030,215.443359)(1.380211,224.183594)(1.447158,231.748047)
(1.477121,235.051758)(1.505150,238.158203)(1.602060,248.828125)(1.681241,257.308594)(1.698970,259.271484)
(1.778151,267.143555)(1.806180,269.520508)(1.845098,272.774414)(1.857332,273.835938)(1.903090,277.277344)
(1.954243,280.917969)(2.000000,283.632812)(2.041393,285.935547)(2.079181,287.601562)(2.113943,289.127930)
(2.146128,290.394531)(2.176091,291.527344)(2.204120,292.517578)(2.230449,293.192383)(2.255273,293.951172)
(2.278754,294.647461)(2.301030,295.106445)(2.322219,295.564453)(2.342423,296.087891)(2.361728,296.654297)
(2.380211,296.879883)(2.397940,297.264648)(2.414973,297.563477)(2.431364,297.779297)(2.447158,298.189453)
(2.462398,298.287109)(2.477121,298.529297)(2.491362,298.643555)(2.505150,298.807617)(2.518514,299.038086)
(2.531479,299.248047)(2.544068,299.356445)(2.602060,300.020508)(2.698970,300.764648)(2.778151,301.080078)
(2.845098,301.242188)(2.903090,300.991211)(2.954243,301.104492)(3.000000,300.947266)(3.041393,300.716797)
(3.079181,300.607422)(3.113943,300.328125)(3.146128,300.303711)(3.176091,300.000977)(3.204120,299.656250)
(3.322219,299.003906)(3.414973,298.377930)(3.491362,297.906250)(3.556303,297.432617)(3.612784,297.142578)
(3.662758,296.888672)

{ \small 
\rput[tl](-1.864094,310.239169){%
\psframebox[framesep=0pt,linewidth=\AxesLineWidth]{\psframebox*{\begin{tabular}{l}
\Rnode{a1}{\hspace*{0.0ex}} \hspace*{0.7cm} \Rnode{a2}{~~BZ2} \\
\Rnode{a3}{\hspace*{0.0ex}} \hspace*{0.7cm} \Rnode{a4}{~~llJP} \\
\Rnode{a5}{\hspace*{0.0ex}} \hspace*{0.7cm} \Rnode{a6}{~~JP2k} \\
\end{tabular}}
\ncline[linestyle=solid,linewidth=\LineWidth,linecolor=color593.0037]{a1}{a2} \ncput{\psdot[dotstyle=Basterisk,dotsize=\MarkerSize,linecolor=color593.0037]}
\ncline[linestyle=solid,linewidth=\LineWidth,linecolor=color594.0027]{a3}{a4} \ncput{\psdot[dotstyle=Bo,dotsize=\MarkerSize,linecolor=color594.0027]}
\ncline[linestyle=solid,linewidth=\LineWidth,linecolor=color595.0027]{a5}{a6} \ncput{\psdot[dotstyle=Btriangle,dotsize=\MarkerSize,linecolor=color595.0027]}
}%
}%
} 

\end{pspicture}%
%
%
%
\providelength{\AxesLineWidth}       \setlength{\AxesLineWidth}{0.5pt}%
\providelength{\plotwidth}           \setlength{\plotwidth}{4.85cm}
\providelength{\LineWidth}           \setlength{\LineWidth}{0.7pt}%
\providelength{\MarkerSize}          \setlength{\MarkerSize}{4pt}%
\newrgbcolor{GridColor}{0.8 0.8 0.8}%
%
\psset{xunit=0.176592\plotwidth,yunit=0.002465\plotwidth}%
\begin{pspicture}(-2.992442,-58.558388)(3.674434,320.836548)%


\psline[linewidth=\AxesLineWidth,linecolor=GridColor](-2.000000,0.000000)(-2.000000,4.868712)
\psline[linewidth=\AxesLineWidth,linecolor=GridColor](0.000000,0.000000)(0.000000,4.868712)
\psline[linewidth=\AxesLineWidth,linecolor=GridColor](2.000000,0.000000)(2.000000,4.868712)
\psline[linewidth=\AxesLineWidth,linecolor=GridColor](-2.000000,0.000000)(-1.932047,0.000000)
\psline[linewidth=\AxesLineWidth,linecolor=GridColor](-2.000000,100.000000)(-1.932047,100.000000)
\psline[linewidth=\AxesLineWidth,linecolor=GridColor](-2.000000,200.000000)(-1.932047,200.000000)
\psline[linewidth=\AxesLineWidth,linecolor=GridColor](-2.000000,300.000000)(-1.932047,300.000000)


{ \footnotesize 
\rput[t](-2.000000,-4.868712){$10^{-2}$}
\rput[t](0.000000,-4.868712){$10^{0}$}
\rput[t](2.000000,-4.868712){$10^{2}$}
\rput[r](-2.067953,0.000000){$0$}
\rput[r](-2.067953,100.000000){$100$}
\rput[r](-2.067953,200.000000){$200$}
\rput[r](-2.067953,300.000000){$300$}
} 

\psframe[linewidth=\AxesLineWidth,dimen=middle](-2.000000,0.000000)(3.662758,320.000000)

{ \small 
\rput[b](0.831379,-58.558388){
\begin{tabular}{c}
noise standard deviation $\sigma$\\
\end{tabular}
}

\rput[t]{90}(-2.992442,160.000000){
\begin{tabular}{c}
file size in kbyte\\
\end{tabular}
}
} 

\newrgbcolor{color593.0012}{0  1  0}
\psline[plotstyle=line,linejoin=1,showpoints=true,dotstyle=Basterisk,dotsize=\MarkerSize,linestyle=solid,linewidth=\LineWidth,linecolor=color593.0012]
(-2.000000,59.809570)(-1.301030,59.944336)(-1.221849,59.817383)(-1.154902,59.816406)(-1.096910,59.814453)
(-1.045757,59.820312)(-1.000000,59.917969)(-0.698970,63.319336)(-0.522879,75.601562)(-0.397940,86.024414)
(-0.301030,92.497070)(-0.221849,97.791992)(-0.154902,101.832031)(-0.096910,105.573242)(-0.045757,108.254883)
(0.000000,111.579102)(0.301030,133.830078)(0.602060,157.813477)(0.903090,183.429688)(1.000000,192.084961)
(1.079181,198.392578)(1.204120,207.844727)(1.301030,214.527344)(1.380211,219.212891)(1.447158,222.379883)
(1.477121,223.708008)(1.505150,224.994141)(1.602060,228.480469)(1.681241,230.148438)(1.698970,230.354492)
(1.778151,230.354492)(1.806180,230.441406)(1.845098,229.855469)(1.857332,229.738281)(1.903090,228.493164)
(1.954243,227.195312)(2.000000,224.490234)(2.041393,222.093750)(2.079181,218.883789)(2.113943,215.803711)
(2.146128,212.010742)(2.176091,208.318359)(2.204120,204.638672)(2.230449,200.850586)(2.255273,197.372070)
(2.278754,194.308594)(2.301030,190.152344)(2.322219,186.879883)(2.342423,183.888672)(2.361728,180.720703)
(2.380211,177.431641)(2.397940,174.406250)(2.414973,171.244141)(2.431364,168.309570)(2.447158,165.539062)
(2.462398,163.710938)(2.477121,160.685547)(2.491362,158.175781)(2.505150,155.615234)(2.518514,153.472656)
(2.531479,151.480469)(2.544068,149.166992)(2.602060,140.803711)(2.698970,125.506836)(2.778151,115.514648)
(2.845098,107.334961)(2.903090,101.629883)(2.954243,95.476562)(3.000000,91.442383)(3.041393,87.815430)
(3.079181,85.171875)(3.113943,82.404297)(3.146128,79.126953)(3.176091,77.199219)(3.204120,75.665039)
(3.322219,68.928711)(3.414973,64.185547)(3.491362,61.248047)(3.556303,58.800781)(3.612784,56.693359)
(3.662758,55.562500)

\newrgbcolor{color594.0002}{1  0  1}
\psline[plotstyle=line,linejoin=1,showpoints=true,dotstyle=Bo,dotsize=\MarkerSize,linestyle=solid,linewidth=\LineWidth,linecolor=color594.0002]
(-2.000000,72.696289)(-1.301030,72.696289)(-1.221849,72.696289)(-1.154902,72.696289)(-1.096910,72.696289)
(-1.045757,72.696289)(-1.000000,72.696289)(-0.698970,73.712891)(-0.522879,80.151367)(-0.397940,88.271484)
(-0.301030,94.851562)(-0.221849,98.131836)(-0.154902,101.625977)(-0.096910,105.465820)(-0.045757,108.196289)
(0.000000,110.636719)(0.301030,130.257812)(0.602060,155.078125)(0.903090,183.351562)(1.000000,192.382812)
(1.079181,201.053711)(1.204120,213.296875)(1.301030,221.495117)(1.380211,229.273438)(1.447158,235.047852)
(1.477121,237.402344)(1.505150,239.327148)(1.602060,247.544922)(1.681241,252.506836)(1.698970,253.555664)
(1.778151,258.333008)(1.806180,259.781250)(1.845098,261.076172)(1.857332,261.298828)(1.903090,262.846680)
(1.954243,264.984375)(2.000000,265.910156)(2.041393,266.023438)(2.079181,266.178711)(2.113943,266.463867)
(2.146128,265.708984)(2.176091,264.643555)(2.204120,263.825195)(2.230449,262.864258)(2.255273,262.233398)
(2.278754,261.197266)(2.301030,260.556641)(2.322219,259.513672)(2.342423,259.062500)(2.361728,257.539062)
(2.380211,255.668945)(2.397940,253.980469)(2.414973,252.519531)(2.431364,251.571289)(2.447158,249.980469)
(2.462398,248.973633)(2.477121,247.593750)(2.491362,246.002930)(2.505150,245.104492)(2.518514,243.767578)
(2.531479,242.555664)(2.544068,241.352539)(2.602060,237.416016)(2.698970,228.933594)(2.778151,223.680664)
(2.845098,219.304688)(2.903090,215.966797)(2.954243,212.826172)(3.000000,210.959961)(3.041393,208.627930)
(3.079181,207.728516)(3.113943,206.298828)(3.146128,204.605469)(3.176091,203.439453)(3.204120,202.775391)
(3.322219,199.472656)(3.414973,197.265625)(3.491362,195.779297)(3.556303,195.036133)(3.612784,193.825195)
(3.662758,192.726562)

\newrgbcolor{color595.0002}{0  0  1}
\psline[plotstyle=line,linejoin=1,showpoints=true,dotstyle=Btriangle,dotsize=\MarkerSize,linestyle=solid,linewidth=\LineWidth,linecolor=color595.0002]
(-2.000000,45.942383)(-1.301030,45.942383)(-1.221849,45.942383)(-1.154902,45.942383)(-1.096910,45.942383)
(-1.045757,45.942383)(-1.000000,45.942383)(-0.698970,48.366211)(-0.522879,59.509766)(-0.397940,69.283203)
(-0.301030,75.127930)(-0.221849,79.118164)(-0.154902,82.548828)(-0.096910,85.755859)(-0.045757,88.797852)
(0.000000,91.628906)(0.301030,115.006836)(0.602060,142.964844)(0.903090,173.194336)(1.000000,183.287109)
(1.079181,191.290039)(1.204120,203.958008)(1.301030,213.390625)(1.380211,221.216797)(1.447158,227.404297)
(1.477121,230.243164)(1.505150,232.824219)(1.602060,241.795898)(1.681241,248.770508)(1.698970,250.450195)
(1.778151,257.362305)(1.806180,259.944336)(1.845098,263.293945)(1.857332,264.250000)(1.903090,268.247070)
(1.954243,272.518555)(2.000000,275.824219)(2.041393,278.808594)(2.079181,281.514648)(2.113943,283.740234)
(2.146128,285.671875)(2.176091,287.158203)(2.204120,288.569336)(2.230449,289.678711)(2.255273,290.872070)
(2.278754,291.555664)(2.301030,292.500977)(2.322219,293.147461)(2.342423,293.962891)(2.361728,294.595703)
(2.380211,295.092773)(2.397940,295.510742)(2.414973,296.013672)(2.431364,296.383789)(2.447158,296.839844)
(2.462398,297.034180)(2.477121,297.285156)(2.491362,297.731445)(2.505150,297.909180)(2.518514,298.121094)
(2.531479,298.329102)(2.544068,298.598633)(2.602060,299.322266)(2.698970,300.238281)(2.778151,300.912109)
(2.845098,300.957031)(2.903090,300.996094)(2.954243,300.983398)(3.000000,300.886719)(3.041393,300.688477)
(3.079181,300.663086)(3.113943,300.594727)(3.146128,300.207031)(3.176091,299.916016)(3.204120,299.837891)
(3.322219,299.045898)(3.414973,298.521484)(3.491362,298.074219)(3.556303,297.511719)(3.612784,297.037109)
(3.662758,296.778320)

{ \small 
\rput[tl](-1.864094,310.262577){%
\psframebox[framesep=0pt,linewidth=\AxesLineWidth]{\psframebox*{\begin{tabular}{l}
\Rnode{a1}{\hspace*{0.0ex}} \hspace*{0.7cm} \Rnode{a2}{~~BZ2} \\
\Rnode{a3}{\hspace*{0.0ex}} \hspace*{0.7cm} \Rnode{a4}{~~llJP} \\
\Rnode{a5}{\hspace*{0.0ex}} \hspace*{0.7cm} \Rnode{a6}{~~JP2k} \\
\end{tabular}}
\ncline[linestyle=solid,linewidth=\LineWidth,linecolor=color593.0012]{a1}{a2} \ncput{\psdot[dotstyle=Basterisk,dotsize=\MarkerSize,linecolor=color593.0012]}
\ncline[linestyle=solid,linewidth=\LineWidth,linecolor=color594.0002]{a3}{a4} \ncput{\psdot[dotstyle=Bo,dotsize=\MarkerSize,linecolor=color594.0002]}
\ncline[linestyle=solid,linewidth=\LineWidth,linecolor=color595.0002]{a5}{a6} \ncput{\psdot[dotstyle=Btriangle,dotsize=\MarkerSize,linecolor=color595.0002]}
}%
}%
} 

\end{pspicture}%
%
%
%
\providelength{\AxesLineWidth}       \setlength{\AxesLineWidth}{0.5pt}%
\providelength{\plotwidth}           \setlength{\plotwidth}{4.85cm}
\providelength{\LineWidth}           \setlength{\LineWidth}{0.7pt}%
\providelength{\MarkerSize}          \setlength{\MarkerSize}{4pt}%
\newrgbcolor{GridColor}{0.8 0.8 0.8}%
%
\psset{xunit=0.176592\plotwidth,yunit=0.002465\plotwidth}%
\begin{pspicture}(-2.992442,-58.558388)(3.674434,320.836548)%


\psline[linewidth=\AxesLineWidth,linecolor=GridColor](-2.000000,0.000000)(-2.000000,4.868712)
\psline[linewidth=\AxesLineWidth,linecolor=GridColor](0.000000,0.000000)(0.000000,4.868712)
\psline[linewidth=\AxesLineWidth,linecolor=GridColor](2.000000,0.000000)(2.000000,4.868712)
\psline[linewidth=\AxesLineWidth,linecolor=GridColor](-2.000000,0.000000)(-1.932047,0.000000)
\psline[linewidth=\AxesLineWidth,linecolor=GridColor](-2.000000,100.000000)(-1.932047,100.000000)
\psline[linewidth=\AxesLineWidth,linecolor=GridColor](-2.000000,200.000000)(-1.932047,200.000000)
\psline[linewidth=\AxesLineWidth,linecolor=GridColor](-2.000000,300.000000)(-1.932047,300.000000)


{ \footnotesize 
\rput[t](-2.000000,-4.868712){$10^{-2}$}
\rput[t](0.000000,-4.868712){$10^{0}$}
\rput[t](2.000000,-4.868712){$10^{2}$}
\rput[r](-2.067953,0.000000){$0$}
\rput[r](-2.067953,100.000000){$100$}
\rput[r](-2.067953,200.000000){$200$}
\rput[r](-2.067953,300.000000){$300$}
} 

\psframe[linewidth=\AxesLineWidth,dimen=middle](-2.000000,0.000000)(3.662758,320.000000)

{ \small 
\rput[b](0.831379,-58.558388){
\begin{tabular}{c}
noise standard deviation $\sigma$\\
\end{tabular}
}

\rput[t]{90}(-2.992442,160.000000){
\begin{tabular}{c}
file size in kbyte\\
\end{tabular}
}
} 

\newrgbcolor{color593.0024}{0  1  0}
\psline[plotstyle=line,linejoin=1,showpoints=true,dotstyle=Basterisk,dotsize=\MarkerSize,linestyle=solid,linewidth=\LineWidth,linecolor=color593.0024]
(-2.000000,132.562500)(-1.301030,132.504883)(-1.221849,132.560547)(-1.154902,132.555664)(-1.096910,132.552734)
(-1.045757,132.571289)(-1.000000,132.470703)(-0.698970,133.204102)(-0.522879,137.103516)(-0.397940,139.892578)
(-0.301030,141.888672)(-0.221849,143.651367)(-0.154902,145.324219)(-0.096910,147.003906)(-0.045757,147.974609)
(0.000000,149.312500)(0.301030,160.102539)(0.602060,174.765625)(0.903090,193.994141)(1.000000,200.991211)
(1.079181,207.266602)(1.204120,217.992188)(1.301030,226.628906)(1.380211,233.493164)(1.447158,239.107422)
(1.477121,241.232422)(1.505150,243.463867)(1.602060,250.079102)(1.681241,253.769531)(1.698970,254.385742)
(1.778151,256.142578)(1.806180,256.255859)(1.845098,256.030273)(1.857332,255.966797)(1.903090,254.907227)
(1.954243,252.228516)(2.000000,248.849609)(2.041393,244.717773)(2.079181,239.623047)(2.113943,234.687500)
(2.146128,229.682617)(2.176091,224.333984)(2.204120,219.010742)(2.230449,214.363281)(2.255273,209.530273)
(2.278754,204.196289)(2.301030,200.060547)(2.322219,195.773438)(2.342423,191.909180)(2.361728,187.612305)
(2.380211,184.096680)(2.397940,180.302734)(2.414973,177.059570)(2.431364,174.079102)(2.447158,170.646484)
(2.462398,167.628906)(2.477121,164.425781)(2.491362,162.027344)(2.505150,159.877930)(2.518514,156.645508)
(2.531479,154.571289)(2.544068,152.191406)(2.602060,142.122070)(2.698970,127.047852)(2.778151,116.783203)
(2.845098,108.216797)(2.903090,101.375977)(2.954243,96.289062)(3.000000,91.741211)(3.041393,87.876953)
(3.079181,84.807617)(3.113943,82.241211)(3.146128,79.635742)(3.176091,77.326172)(3.204120,75.833984)
(3.322219,69.234375)(3.414973,64.675781)(3.491362,61.377930)(3.556303,58.905273)(3.612784,57.129883)
(3.662758,55.701172)

\newrgbcolor{color594.0015}{1  0  1}
\psline[plotstyle=line,linejoin=1,showpoints=true,dotstyle=Bo,dotsize=\MarkerSize,linestyle=solid,linewidth=\LineWidth,linecolor=color594.0015]
(-2.000000,141.943359)(-1.301030,141.943359)(-1.221849,141.943359)(-1.154902,141.943359)(-1.096910,141.943359)
(-1.045757,141.943359)(-1.000000,141.943359)(-0.698970,142.073242)(-0.522879,142.673828)(-0.397940,143.198242)
(-0.301030,143.745117)(-0.221849,144.329102)(-0.154902,144.885742)(-0.096910,145.477539)(-0.045757,146.120117)
(0.000000,146.852539)(0.301030,153.781250)(0.602060,167.116211)(0.903090,188.068359)(1.000000,196.815430)
(1.079181,204.715820)(1.204120,215.875000)(1.301030,225.542969)(1.380211,234.084961)(1.447158,241.097656)
(1.477121,244.037109)(1.505150,246.498047)(1.602060,256.140625)(1.681241,264.608398)(1.698970,266.241211)
(1.778151,273.058594)(1.806180,275.302734)(1.845098,277.895508)(1.857332,278.484375)(1.903090,280.341797)
(1.954243,282.192383)(2.000000,283.235352)(2.041393,283.646484)(2.079181,283.506836)(2.113943,283.072266)
(2.146128,283.037109)(2.176091,281.310547)(2.204120,279.482422)(2.230449,276.314453)(2.255273,274.276367)
(2.278754,272.674805)(2.301030,270.321289)(2.322219,268.516602)(2.342423,266.686523)(2.361728,265.052734)
(2.380211,263.434570)(2.397940,262.174805)(2.414973,260.096680)(2.431364,258.472656)(2.447158,256.343750)
(2.462398,254.625000)(2.477121,252.985352)(2.491362,251.092773)(2.505150,250.097656)(2.518514,248.350586)
(2.531479,247.051758)(2.544068,245.955078)(2.602060,240.200195)(2.698970,231.187500)(2.778151,225.496094)
(2.845098,220.327148)(2.903090,216.360352)(2.954243,213.984375)(3.000000,210.979492)(3.041393,209.166016)
(3.079181,207.785156)(3.113943,205.900391)(3.146128,204.647461)(3.176091,203.316406)(3.204120,202.481445)
(3.322219,199.613281)(3.414973,197.033203)(3.491362,195.922852)(3.556303,195.083984)(3.612784,193.893555)
(3.662758,193.499023)

\newrgbcolor{color595.0015}{0  0  1}
\psline[plotstyle=line,linejoin=1,showpoints=true,dotstyle=Btriangle,dotsize=\MarkerSize,linestyle=solid,linewidth=\LineWidth,linecolor=color595.0015]
(-2.000000,85.889648)(-1.301030,85.889648)(-1.221849,85.889648)(-1.154902,85.889648)(-1.096910,85.889648)
(-1.045757,85.889648)(-1.000000,85.889648)(-0.698970,86.688477)(-0.522879,90.668945)(-0.397940,94.585938)
(-0.301030,97.464844)(-0.221849,99.918945)(-0.154902,102.293945)(-0.096910,104.494141)(-0.045757,106.626953)
(0.000000,108.742188)(0.301030,127.117188)(0.602060,150.974609)(0.903090,178.473633)(1.000000,187.921875)
(1.079181,195.649414)(1.204120,208.200195)(1.301030,218.339844)(1.380211,226.505859)(1.447158,233.622070)
(1.477121,236.458008)(1.505150,239.400391)(1.602060,249.434570)(1.681241,257.410156)(1.698970,258.995117)
(1.778151,266.340820)(1.806180,268.848633)(1.845098,271.859375)(1.857332,272.916016)(1.903090,276.233398)
(1.954243,279.918945)(2.000000,282.601562)(2.041393,284.961914)(2.079181,286.872070)(2.113943,288.565430)
(2.146128,289.749023)(2.176091,290.981445)(2.204120,292.032227)(2.230449,292.827148)(2.255273,293.549805)
(2.278754,294.238281)(2.301030,294.916016)(2.322219,295.315430)(2.342423,295.923828)(2.361728,296.375000)
(2.380211,296.778320)(2.397940,297.093750)(2.414973,297.288086)(2.431364,297.699219)(2.447158,298.015625)
(2.462398,298.163086)(2.477121,298.425781)(2.491362,298.590820)(2.505150,298.761719)(2.518514,299.018555)
(2.531479,299.121094)(2.544068,299.269531)(2.602060,299.953125)(2.698970,300.652344)(2.778151,300.979492)
(2.845098,301.126953)(2.903090,301.119141)(2.954243,301.006836)(3.000000,300.790039)(3.041393,300.716797)
(3.079181,300.546875)(3.113943,300.336914)(3.146128,300.167969)(3.176091,299.846680)(3.204120,299.785156)
(3.322219,298.961914)(3.414973,298.351562)(3.491362,297.807617)(3.556303,297.474609)(3.612784,297.015625)
(3.662758,296.661133)

{ \small 
\rput[tl](-1.864094,310.262577){%
\psframebox[framesep=0pt,linewidth=\AxesLineWidth]{\psframebox*{\begin{tabular}{l}
\Rnode{a1}{\hspace*{0.0ex}} \hspace*{0.7cm} \Rnode{a2}{~~BZ2} \\
\Rnode{a3}{\hspace*{0.0ex}} \hspace*{0.7cm} \Rnode{a4}{~~llJP} \\
\Rnode{a5}{\hspace*{0.0ex}} \hspace*{0.7cm} \Rnode{a6}{~~JP2k} \\
\end{tabular}}
\ncline[linestyle=solid,linewidth=\LineWidth,linecolor=color593.0024]{a1}{a2} \ncput{\psdot[dotstyle=Basterisk,dotsize=\MarkerSize,linecolor=color593.0024]}
\ncline[linestyle=solid,linewidth=\LineWidth,linecolor=color594.0015]{a3}{a4} \ncput{\psdot[dotstyle=Bo,dotsize=\MarkerSize,linecolor=color594.0015]}
\ncline[linestyle=solid,linewidth=\LineWidth,linecolor=color595.0015]{a5}{a6} \ncput{\psdot[dotstyle=Btriangle,dotsize=\MarkerSize,linecolor=color595.0015]}
}%
}%
} 

\end{pspicture}%
\vspace*{-1mm}

\hfill{}(a)\hfill{}\hfill{}(b)\hfill{}\hfill{}(c)\hfill{}

\vspace*{-2mm}

\protect\caption{\label{fig:SimulationBitsOverVeriance}Simulation results show the
file size in kbytes over the standard deviation $\sigma$ of the added
noise in lin-log plots. (a) shows the results for a constant signal
equal to 128. In (b) the results for the first test image (Figure~\ref{fig:SimulationTestimages}~(a))
are shown. (c) shows the results for the second test image (Figure~\ref{fig:SimulationTestimages}~(b))\vspace*{-5mm}}
\end{figure*}

In order to prove that our model also fits for images we used details
from the \emph{Big Buck Bunny} sequence. The reason for the choice
of artificial images is the difficulty of perfectly denoising real
images \cite{chatterjee2010}. The chosen images are computer generated
and thus do not contain noise in the beginning. We cut images of 512x512
pixels as shown in Figure~\ref{fig:SimulationTestimages}~(a) and
(b). For illustration we added Gaussian noise with a standard deviation
of $\sigma=30$ to the first detail shown in Figure~\ref{fig:SimulationTestimages}~(a).
The second detail in Figure~\ref{fig:SimulationTestimages}~(b)
is shown in original, i.e., without additive noise. In our simulation
we used the green color channel only.

For \emph{direct} entropy coding we used bzip2 \cite{bzip2} in version~1.0.4.
No additional parameters were given when calling the program. For
\emph{predictive} coding we used lossless~JPEG \cite{losslessJPEG,wallace1992}.
For \emph{wavelet}-based coding we used  JPEG~2000 \cite{jpeg2k,christopoulos2002}
 with 4~decomposition steps. As long as the number of decomposition
steps is larger than 3 this parameter has not much influence on the
file size.

The results are shown in Figure~\ref{fig:SimulationBitsOverVeriance}
where the file size of the compressed images is plotted in kbyte over
the standard deviation $\sigma$ of the noise in lin-log plots.

In Figure~\ref{fig:SimulationBitsOverVeriance}~(a) the results
are shown for the image with a constant code value of 128 and additive
noise. By comparing Figure~\ref{fig:SimulationBitsOverVeriance}~(a)
with the theoretical results in Figure~\ref{fig:TheoryEntropyOverVariance}
it can be seen that our model matches with the results from the simulation
up to the offset of the lossless~JPEG method for small values of
$\sigma$. The corresponding curves have the same shape and the plot
in Figure~\ref{fig:SimulationBitsOverVeriance}~(a) can also be
divided into the four areas indicated in Figure~\ref{fig:TheoryEntropyOverVariance}.

Quantitative statements cannot be given as the results strongly depend
on the structural information in the input signal.

The offsets in Figure~\ref{fig:SimulationBitsOverVeriance}~(b)
and (c) come from the structural information. The three methods need
a different amount of bits to code the structural information. The
two plots show clearly that as long as the noise part is small enough
advanced methods as a \emph{predictive} scheme and a \emph{wavelet}
decomposition are advantageous because they are more capable to reduce
the redundancy in the structural information part of the signal.

Clipping in the structural information, e.g. due to wrong exposure,
often affects bigger areas of an image and thus is beneficial for
\emph{predictive} and \emph{wavelet}-based coding. The problem are
isolated clipped values introduced by noise because their combination
with other unclipped samples lead to a higher entropy.

The results in Figure~\ref{fig:TheoryEntropyOverVariance} and Figure~\ref{fig:SimulationBitsOverVeriance}~(a)
show that it is advantageous to use \emph{direct} entropy coding for
noise. In \cite{chatterjee2010} several state of the art denoising
algorithms are compared to a theoretical bound. The result is that
there is still room for improvements. A separation of the noise from
the structural information leads to an increase of the encoder complexity.
The decoder has to decode both parts and add the noise to the structural
part. Compared with \emph{wavelet}-based coding gains can be achieved
for small values of $\sigma$ as shown in area~(b) of Figure~\ref{fig:TheoryEntropyOverVariance}.

Another result is that for very noisy images when clipping is introduced
by noise it is more advantageous to directly use an entropy coder
without any signal decomposition. These results show that our model
also fits for images with structural information.

\section{Conclusion}

In this paper we analyzed the impact of clipping to lossless compression
of noisy images. We derived an analytical description that models
the behavior of different coding methods. So the effects shown are
general properties of the methods and not of a special implementation.
Simulation results support our model. The results show that in general
for noisy data it is advantageous to code the noise and the signal
separately. Furthermore, the results show that for the case that clipping
is introduced by noise it is more advantageous to directly use an
entropy coder without advanced preprocessing steps.

\section*{Acknowledgment}

We gratefully acknowledge that this work has been supported by the
Deutsche Forschungsgemeinschaft (DFG) under contract number KA~926/4-1.

\end{document}